\documentclass[useAMS,usenatbib]{mn2e}

\include{epsf}
\newcommand{\be}{\begin{equation}}
\newcommand{\ee}{\end{equation}}
\newcommand{\ba}{\begin{eqnarray}}
\newcommand{\ea}{\end{eqnarray}}

\def\simless{\mathbin{\lower 3pt\hbox
   {$\rlap{\raise 4pt\hbox{$\char'074$}}\mathchar"7218$}}}
\def\simgreat{\mathbin{\lower 3pt\hbox
   {$\rlap{\raise 4pt\hbox{$\char'076$}}\mathchar"7218$}}}
\title[SFHs from multiband photometry]
{Star formation histories from multiband photometry: A new approach} 
\author[Simon Dye]{ 
\parbox[t]{\textwidth}{
Simon Dye$^1$\thanks{E-mail: s.dye@astro.cf.ac.uk}
}\\ \\ 
$^1$Cardiff University, School of Physics \& Astronomy, Queens Buildings,
The Parade, Cardiff, CF24 3AA, U.K. \\
}

\begin{document}

\date{Document in prep.}

\pagerange{\pageref{firstpage}--\pageref{lastpage}} \pubyear{2008}

\maketitle

\label{firstpage}

\begin{abstract}

A new method of determining galaxy star-formation histories (SFHs) is
presented. Using the method, the feasibility of recovering SFHs with
multi-band photometry is investigated.  The method divides a galaxy's
history into discrete time intervals and reconstructs the average rate
of star formation in each interval. This directly gives the total
stellar mass.  A simple linear inversion solves the problem of finding
the most likely discretised SFH for a given set of galaxy parameters.
It is shown how formulating the method within a Bayesian framework
lets the data simultaneously select the optimal regularisation
strength and the most appropriate number of discrete time intervals
for the reconstructed SFH. The method is demonstrated by applying it
to mono-metallic synthetic photometric catalogues created with
different input SFHs, assessing how the accuracy of the recovered SFHs
and stellar masses depend on the photometric passband set,
signal-to-noise and redshift. The results show that reconstruction of
SFHs using multi-band photometry is possible, being able to
distinguish an early burst of star formation from a late one, provided
an appropriate passband set is used. Although the resolution of the
recovered SFHs is on average inferior compared to what can be achieved
with spectroscopic data, the multi-band approach can process a
significantly larger number of galaxies per unit exposure time.

\end{abstract}


\section{Introduction}
\label{sec_intro}

A significant step towards understanding how galaxies form and evolve
can be made by measuring the variation in their star formation rate
(SFR) with age.  Imprinted in every galaxy's integrated light is a
record of its entire life from birth, through passive evolution,
possible merging and recycling of material, up to the epoch at which
it is observed. Star formation histories (SFHs) therefore play a
crucial role in the quest for a complete and accurate model of the
formation of stellar mass in the Universe and how distant systems
relate to those locally.

Characterising galaxy SFHs has been a subject of much interest for
several decades, with studies attempting to achieve this aim through a
variety of different means. Approaches can be broadly divided into
those using multi-band photometry and those using spectra. Recently
the practice has seen a significant revival thanks to improvements in
stellar synthesis modelling and the advent of large datasets such as
the Sloan Digital Sky Survey \citep[SDSS;][]{stoughton02}. Many new
spectroscopic techniques have been developed \citep[e.g.,][]{heavens00,
vergely02,cid04,cid05,nolan06,ocvirk06,chil07,tojeiro07} and in their
various forms, these have seen application to several sets of real
data \citep[e.g.,][] {reichardt01,panter03,heavens04,panter04,sheth06,
cid07,nolan07,panter07,koleva08}. Similarly, there have been
numerous recent studies conducted using multi-band photometry
\citep[e.g.,][]{borch06,schawinski07,salim07,noekse07,kaviraj07}
including \citet{kauffmann03} who combined multi-band photometry with
measurements of the H$\delta$ absorption line and 4000\AA break
strength.

In a similar vein to spectroscopic versus photometric redshift
estimation, SFHs determined from spectra tend to have greater
precision per galaxy, whereas those derived from multi-band photometry
allow many more objects to be studied in the same amount of observing
time but with a compromise in SFH resolution.  The method adopted by
existing multi-band studies is to assume a parametric model for the
SFH. The parameters are adjusted to find the set of model fluxes,
computed from a spectral library of choice, that best matches the set
of observed fluxes. This not only forces the SFH to adhere to a
potentially unrepresentative prescribed form, it also necessitates a
fully non-linear minimisation over all parameters.  

In contrast, the majority of the recent spectroscopic methods divide
up a galaxy's history into several independent time intervals and
reconstruct the average SFR in each interval to give a discretised
SFH. The advantage this brings, as shown in Section
\ref{sec_most_prob_SFH}, is that finding the best-fit SFR in every
interval for a fixed set of galaxy parameters (such as redshift,
extinction and metallicity) is a linear problem. The inefficient
non-linear SFH minimisation with its risk of becoming trapped in local
minima is therefore replaced with a simple matrix inversion
guaranteeing that the global minimum for the fixed set of galaxy
parameters is found.

The prescribed SFH models used by the multi-band methods are mainly
driven by the small number of passbands used in many multi-band
campaigns. With only a small number of passbands, the ability to
constrain a galaxy's SFH is limited and a model SFH with only one or
two parameters must be used. However, modern surveys are being carried
out in many more passbands and over larger wavelength ranges than ever
before \citep[for example, the COMBO-17 survey of][]{wolf01}. Given
these recent improvements, the possibility of recovering discretised
SFHs from multi-band photometry alone is now worthy of investigation.

The purpose of this paper is twofold. Firstly, a new SFH
reconstruction method that recovers discretised SFHs is presented. It
is shown how the Bayesian evidence can be used to simultaneously
establish the most appropriate number of discrete SFH time intervals
and the optimal strength with which the solution should be
regularised. The formalism is completely general and can be applied to
spectra just as easily as multi-band photometry as well as a
combination of both. The Bayesian evidence gives a more natural and
simplified alternative to existing procedures for determining the
optimal number of SFH intervals and for determining the correct level
of regularisation.

Secondly, this paper presents results of an investigation into the
feasibility of using the new method with multi-band photometry alone.
By applying the method to synthetic galaxy catalogues created with
different input SFHs and filtersets, the accuracy of the recovered
discretised SFHs is demonstrated. This study focuses in particular on
the dependence of the reconstruction on galaxy redshift, photometric
signal-to-noise (S/N), the wavelength range spanned by the passbands,
the number of passbands and the presence/absence of a new and/or old
stellar population.

The layout of the paper is as follows. In Section \ref{sec_method}
the SFH reconstruction method is described. Section
\ref{sec_synthetic_cats} gives details of how the synthetic catalogues
are generated. The method is applied to these catalogues in Section
\ref{sec_sims} to assess its performance.  Section \ref{sec_summary}
gives a summary of the findings of this paper to act as recommendations
for applying the method to real data.

Throughout this paper, the following cosmological parameters are
assumed; ${\rm H}_0=100\,{\rm h}_0=70\,{\rm km\,s}^{-1}\,{\rm
Mpc}^{-1}$, $\Omega_m=0.3$, $\Omega_{\Lambda}=0.7$. All magnitudes are
expressed in the AB system.

\section{The method}
\label{sec_method}

The method divides a galaxy's history into discrete blocks
of time. The goal is to establish the average star formation rate
(SFR) in each block to arrive at a discretised SFH that best fits the
observed galaxy multi-band photometry. As shown in Section
\ref{sec_sims}, the optimal number of blocks is a function of many
attributes, including the number of filters in which the galaxy has
been observed and the signal-to-noise (S/N) of the data.

\subsection{Determination of model fluxes}
\label{sec_model_fluxes}

In order to proceed, a model flux must be determined in each passband
from the discretised SFH to establish the goodness of fit with the
observed fluxes.  For the purpose of demonstration, in this paper, the
synthetic spectral libraries of \citet{bruzual03} are used to compute
the SED for each SFH block although the method is completely general
and can be applied with any empirical or synthetic library.

Starting with a simple stellar population (SSP) SED, $L_{\lambda}^{\rm
SSP}$, of metallicity $Z$, a composite stellar population (CSP) SED,
$L_{\lambda}^{\, i}$, is generated for the $i$th block of constant
star formation in a given galaxy using
\be
L_{\lambda}^{\, i} = \frac{1}{\Delta t_i} \int^{t_i}_{t_{i-1}} {\rm d}t' \, 
L_{\lambda}^{\rm SSP}(\tau(z)-t')
\ee
where the block spans the period $t_{i-1}$ to $t_i$ in the galaxy's
history and $\tau$ is the age of the galaxy (i.e., the age of the
Universe today minus the look-back time to the galaxy). The
normalisation $\Delta t_i=t_i-t_{i-1}$ ensures that the CSP has the
same normalisation as the SSP which in the case of the
\citet{bruzual03} libraries is one solar mass. In practice, the
integration is replaced by a sum over the SSP SEDs which are defined
at discrete time intervals.  In the present work, this sum is carried
out over finer intervals than the library provides by interpolating
the SSP SEDs linearly in log($t$).  Note also that this work considers
mono-metallic stellar populations such that $Z$ does not vary with
age. The more general problem of allowing $Z$ to evolve with time is
left for future work (see Section \ref{sec_summary}).
 
To model the effects of extinction on the final SED (i.e., the SED
from all blocks in the SFH), reddening is applied. This is achieved by
individually reddening the CSP of each block using
\be
L_{\lambda,R}^{\, i} = L_{\lambda}^{\, i} \,
10^{-0.4 k(\lambda)A_V/R_V} \, .
\ee
Here, $A_V$ is the extinction and $k(\lambda)$ is taken as the 
Calzetti law for starbursts \citep{calzetti00},
\be
k(\lambda) = 
\left\{ \begin{array}{l}
2.659(-2.156+\frac{1.509}{\lambda}-\frac{0.198}{\lambda^2}+
\frac{0.011}{\lambda^3})+R_V \\
\hspace{15mm} ({\rm for}\,\,\, 0.12\mu{\rm m} < \lambda <  0.63\mu{\rm m}) \\
2.659(-1.857+\frac{1.04}{\lambda})+R_V \\
\hspace{15mm}  ({\rm for \,\,\,} 0.63\mu{\rm m} < \lambda <  2.2\mu{\rm m}) 
\end{array} \right .
\ee 
with $R_V=4.05$ and $\lambda$ in microns.
To match the wavelength range of the passbands considered in
this study, it is assumed that the longer wavelength half of the
function applies up to 10$\mu$m and the shorter wavelength half is
extrapolated down to 0.01$\mu$m using the average slope between
0.12$\mu$m and 0.13$\mu$m.  The model flux (i.e., photon count)
observed in passband $j$ from a given block $i$ in the SFH when the
galaxy lies at a redshift $z$ is then
\be
F_{ij}=\frac{1}{4\pi d_L^{\,2}}\int {\rm d}\lambda 
\frac{\lambda \, L^i_{\lambda,R}(\lambda/(1+z))T_j(\lambda)}{(1+z)\, hc} 
\ee
where $d_L$ is the luminosity distance and $T_j$ is the transmission
curve of passband $j$. 

\subsection{Determination of the most probable SFH}
\label{sec_most_prob_SFH}

To find the normalisations $a_i$ which result in a set of model fluxes
that best fits the observed fluxes, the following $\chi^2$ function
is minimised
\be
\label{eq_chi_sq}
\chi^2=\sum_j^{N_{\rm filt}} \frac{(\sum_i^{N_{\rm block}} \,
a_i F_{ij} - F^{\rm obs}_j)^2}{\sigma_j^2}
\ee
where $F^{\rm obs}_j$ is the flux observed in passband $j$ from the
galaxy and $\sigma_j$ is its error. The sum in $i$ acts over all
$N_{\rm block}$ SFH blocks. In the case of application to
spectroscopic data instead of multi-band photometry, the index $j$
would refer to spectral elements rather than passbands.  $F_{ij}$
would represent the flux of the model SED over the wavelength range
$\lambda_j$ to $\lambda_j+\Delta\lambda$ from SFH block $i$ and
$F^{\rm obs}_j$ would be the corresponding flux from the observed SED.
In fact, the generality of this approach means that
a combination of spectroscopic data and multi-band 
photometry can be used\footnote{In the case of covariant data,
equation (\ref{eq_chi_sq}) would be replaced by the more general
form $\chi^2=\sum_{ij}\,(x_i-y_i)\sigma_{ij}^{-1}(x_j-y_j)$, with
$x_j=\sum_i\,a_i F_{ij}$, $y_j=F^{\rm obs}_j$ and where 
$\sigma_{ij}^{-1}$ is the inverse covariance matrix.}, the 
appropriate weighting being applied by
$\sigma_j^2$. In any case, the total stellar mass of the galaxy
is simply the sum of the mass normalisations of each block:
\be
\label{eq_stellar_mass}
{\rm M}_* = \sum_i^{N_{\rm block}} a_i \, .
\ee

The minimum $\chi^2$ occurs when the condition
$\partial\chi^2/\partial\,a_i=0$ is simultaneously satisfied for all
$a_i$. This is a linear problem with the following solution:
\be
\label{eq_matrix}
\mathbf{a}=\mathbf{G}^{-1}\mathbf{d} \, .
\ee
Here $\mathbf{a}$ is a column vector composed of
the normalisations $a_i$, $\mathbf{G}$ is a $N_{\rm block} 
\times N_{\rm block}$ square matrix whose $ik$th element is 
given by
\be
G_{ik}=\sum_{j=1}^{N_{\rm filt}} \, F_{ij}F_{kj}/\sigma_j^2
\ee
and $\mathbf{d}$ is a one dimensional vector with elements
\be
d_i=\sum_{j=1}^{N_{\rm filt}} \, F_{ij}F_j^{\rm obs}/\sigma_j^2 \, .
\ee
However, in the presence of noise, the solution given by 
equation (\ref{eq_matrix}) is formally ill-conditioned. This
is circumvented by linear regularisation which involves
adding an extra term, the regularisation matrix $\mathbf{H}$,
weighted by the regularisation weight, $w$ (see
Section \ref{sec_regularisation}):
\be
\label{eq_matrix_reg}
\mathbf{a}=(\mathbf{G}+w\mathbf{H})^{-1}\mathbf{d} \,.
\ee
The errors on the normalisations $a_i$ are obtained from
the corresponding covariance matrix which was derived by
\citet{warren03} for this problem:
\be
\label{eq_sfr_errors}
\mathbf{C} = \mathbf{R} - w \mathbf{R}
(\mathbf{R}\mathbf{H})^T
\ee
where the definition $\mathbf{R}=(\mathbf{G}+w\mathbf{H})^{-1}$
has been made for simplicity.

Unfortunately, by regularising the solution, a new problem is
introduced. The effect of regularisation is to reduce the effective
number of degrees of freedom by an amount that can not be
satisfactorily determined. Furthermore, applying the same
regularisation weight to two different models (for example different
numbers of SFH blocks) results in a different effective number of
degrees of freedom for each model \citep{dye05,dye07}. This means the
minimum $\chi^2$ is biased away from the most probable solution. More
crucially, comparison between different models cannot be carried out
fairly using the $\chi^2$ statistic. For example, $\chi^2$ could not
be used to identify the spectral library that best fits a set of
observed fluxes from a selection of libraries.  This characteristic
has been ignored in previous studies.

One solution to the problem is to simply not regularise. Fortunately,
a better solution can be found by turning to Bayesian inference and
ranking models by their Bayesian evidence instead of $\chi^2$ (see
Appendix \ref{sec_app_evidence}). \citet{suyu06} derived an expression
for the Bayesian evidence, $\epsilon$, for the linear inversion
problem described by equation (\ref{eq_matrix_reg}). Using the
previous notation, this can be written
\ba
\label{eq_evidence}
-2 \,{\rm ln} \, \epsilon &=& 
\chi^2 -{\rm ln} \, \left[ {\rm det} (w\mathbf{H})\right]
+{\rm ln} \, \left[ {\rm det} (\mathbf{G}+w\mathbf{H})\right]
\nonumber \\
& & + \, w\mathbf{a}^T\mathbf{H\,a} +
\sum_{j=1}^{N_{\rm filt}}{\rm ln} (2\pi \sigma_j^2) \, 
\ea
with $\chi^2$ given by equation (\ref{eq_chi_sq}).  Here, 
the covariance between all pairs of observed fluxes has been set to
zero (i.e., it is assumed all fluxes are independent of each other. For
covariant data, the more general form given by \citet{suyu06} would
be used).

The evidence is a probability distribution in the model parameters and
regularisation weight, $w$, allowing different models to be ranked
fairly to find the most probable model. Formally, the evidence should
be marginalised over $w$ and the result used in the ranking. However,
\citet{suyu06} noted that the distribution function for $w$ can be
approximated as a delta function centred on the optimal regularisation
weight, $\hat{w}$. This is a reasonable simplification since $\hat{w}$
is a distinct value estimable from the data. With this simplification,
the maximised value of the evidence at $\hat{w}$ can be directly used
to rank models rather than having to maximise the more computationally
demanding marginalised evidence (see Appendix \ref{sec_app_evidence}).
This approximation has been adopted in the present study.

\subsection{Maximisation procedure}
\label{sec_max_proc}

The complete process of establishing the most probable SFH when the
galaxy's redshift, extinction and metallicity ($z$, $A_V$, $Z$) are
unknown is most conveniently separated into three nested 
levels of inference \citep[e.g., see the general approach
to Bayesian inference by][]{mackay03}:
\begin{itemize}

\item[$\bullet$] In the innermost level, the most likely SFH for a given
$z$, $A_V$, $Z$ and number of SFH blocks, $N_{\rm block}$, as well as
a given regularisation weight, $w$, is determined with the
linear inversion step outlined in the previous section.

\item[$\bullet$] In the second level, the most probable $w$ is
determined for a given $z$, $A_V$, $Z$ and $N_{\rm block}$ by
maximising the evidence given in equation (\ref{eq_evidence}).
Quantitatively, this means that equation (\ref{eq_matrix_reg}) must be
evaluated every time $w$ is varied in the evidence maximisation.

\item[$\bullet$] Finally, in the third and outermost level, the set of
parameters $z$, $A_V$, $Z$ and $N_{\rm block}$ which maximise the
evidence from the second level are found.

\end{itemize}

In this paper, a more specific case is considered where $z$ and $A_V$
are known for each galaxy\footnote{Note that assuming prior knowledge
of the extinction is not equivalent to setting $A_V=0$ for all sources
and ignoring it in the maximisation. Assigning non-zero extinction,
despite not being maximised, allows proper exploration of any
systematics or SED degeneracies that might exist.}.  Such a scenario
might arise, for example, if these parameters have been provided
without spectroscopic data, if a spectrum is available but over a
wavelength range too narrow to obtain a reliable SFH, or if the
parameters are known globally for a group or cluster of galaxies but
SFHs are required for individual galaxies. In addition, mono-metallic
stellar populations are considered in this work, such that $Z$ remains
constant at all times throughout the galaxy's history (see Section
\ref{sec_summary} for a discussion of the more general problem). With
these constraints, the third level of inference therefore requires
varying only $N_{\rm block}$ and $Z$.

The reason for segregating $w$ into a separate second level of
inference, rather than combining it with $z$, $A_V$, $Z$ and $N_{\rm
block}$ is twofold. Firstly, it is not a formal parameter of the
fit. Its optimal value is an indication of the quantity of information
the data contain. In Bayesian terms, regularisation takes the role of
a prior since it corresponds to an a priori assumption regarding the
smoothness of the solution (see Appendix \ref{sec_app_evidence}).
Secondly, there is a practical consideration. As \citet{dye08}
discuss, the value of $w$ that maximises the evidence in the
second level of inference varies only slightly with different trial
sets of model parameters in the third level. This means that one can
alternate between varying $w$ whilst fixing $N_{\rm block}$ \&
$Z$ and varying $N_{\rm block}$ \& $Z$ whilst fixing $w$.
Alternating between two separate levels in this way increases the
efficiency of the maximisation.  Furthermore, by starting the
maximisation with $w$ held fixed at a large value, the evidence
varies more smoothly with $N_{\rm block}$ \& $Z$. This gives an
additional improvement in the speed with which the global maximum can
be found and reduces the risk of becoming stuck at local maxima. In
this paper, the alternating maximisation method described is applied,
stopping once the evidence has converged.

In principle, as many parameters as desired can be added in the second
step above. For example, one might like the duration of some or all of
the SFH blocks to vary. Of course, the limiting factor is ultimately
the number of photometric data points. Adding more parameters in the
second stage results in the evidence being maximised at lower SFH
resolutions.  Therefore, to maximise SFH resolution, spacing is kept
fixed in this work. SFH blocks are assigned a duration $c\,b^{-i}$,
where $i$ is the block number (increasing with age), $c$ is a stretch
factor, always set to make the end of the last SFH block coincide with
the age of the galaxy and the parameter $b$ is set to 1.5. This
exponential spacing allocates smaller periods at later times to
account for the fact that a galaxy's SED is more strongly influenced
by more recent star formation activity.

When finding the most probable value of $w$, the downhill simplex
method is used, minimising the quantity $-{\rm ln}
\,\epsilon$. However, $N_{\rm block}$ is a discontinuous parameter
hence to find the most probable $N_{\rm block}$, the evidence is
computed across a range of values of $N_{\rm block}$ and that which
maximises the evidence is selected. In this way, the optimal number of
SFH blocks are automatically selected by the data. Maximising the
evidence is a more natural and simplified alternative to the iterative
procedure used by \citet{tojeiro07} for determining the optimal number
of SFH intervals.  This also simplifies the method used by
\citet{ocvirk06} for determining the level of regularisation.

On a 3 GHz desktop computer, the full process of determining the
regularisation weight, metallicity and number of SFH bursts that
simultaneously maximise the evidence takes approximately three to four
seconds per galaxy for the largest filterset considered in this work
comprising 13 filters (see Section \ref{sec_synthetic_cats}).

\subsection{Regularisation}
\label{sec_regularisation}

In a Bayesian framework, regularisation takes the role of a prior by
assuming a smooth SFH. The effect is to smear out noisy spikes in the
solution. A downside is that real bursts that occur on a short
timescale are also smeared. However, the goal of adopting a relatively
coarsely binned SFH is to recover longer timescale events, aiming for
reliability rather than a high SFH resolution. Furthermore,
regularisation is necessary to ensure that the linear solution given
by equation (\ref{eq_matrix}) is well defined.

Regularisation is achieved by adding an extra term to $\chi^2$ so that
the figure of merit becomes $\chi^2+B$.  Generally, if this term can
be written 
\be
\label{eq_reg_term}
B=\sum_{i,k}\,b_{ik}a_i\,a_k
\ee
where the $b_{ik}$ are constants, then the solution remains linear
since its partial derivative with respect to all normalisations $a_i$
is linear in $a$. The elements of the regularisation matrix
$\mathbf{H}$ introduced in Section \ref{sec_most_prob_SFH}
are related to the regularisation term $B$ via
\be
2\,H_{ik}=\frac{\partial^2 B}{\partial a_i \partial a_k} \, .
\ee

The most basic form of regularisation, known as {\em zeroth order}
regularisation, is obtained by setting $b_{ik}=\delta_{ik}$. In this
case, the regularisation term to be minimised becomes
$B=\sum_i\,a_i^2$. In {\em first order} regularisation, the
regularisation term is written $B=\sum_i\,(a_i-a_{i+1})^2$ and for
{\em second order}, $B=\sum_i\,(2a_i-a_{i-1}-a_{i+1})^2$. In
principle, the most appropriate type of regularisation to apply can be
decided by the Bayesian evidence.  However, in this work, for
simplicity and to keep the number of non-linear parameters to a
minimum, the regularisation type was fixed. Zeroth-order
regularisation was rejected on the grounds that it prefers
non-physical, null SFH solutions. Tests revealed that 
second order regularisation results in slightly more accurate
SFHs than first order on average, hence second order was applied
to all reconstructions in this paper.

A final consideration regarding regularisation is that the matrix
$\mathbf{H}$ must not be singular. Ensuring the non-singularity of
$\mathbf{H}$ ensures that the evidence, which depends on 
ln[det($\mathbf{H}$)], can be calculated. To guarantee non-singularity,
the following was used for the regularisation term: 
\ba
B&=&(a_{N_{\rm block}}-a_{N_{\rm block}-1})^2 + (a_1-a_2)^2\nonumber \\
&+& \sum_{i=2}^{N_{\rm block}-1}\,(2a_i-a_{i-1}-a_{i+1})^2 \, .
\ea

\section{Synthetic catalogues}
\label{sec_synthetic_cats}

The performance of the SFH reconstruction method was tested by
applying it to a suite of different synthetic galaxy catalogues.  The
suite was designed to encompass a range of SFHs and filter
sets for assessing how the recovered SFH and total stellar mass
depends on each permutation. All catalogues were constructed using 
\citet{bruzual03} SED libraries with the 1994 Padova evolutionary
tracks \citep{bertelli94} and Salpeter initial mass function
\citep{salpeter55} using the method outlined in Section
\ref{sec_model_fluxes}.  Although the exact numerical results will 
depend on the library used, the observed global trends would be
expected to hold true generally.

\begin{figure}
\epsfxsize=8cm
{\hfill
\epsfbox{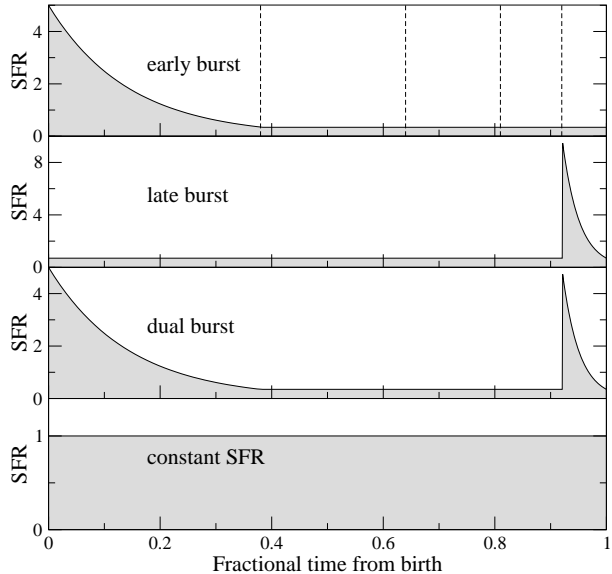}
\hfill}
\epsfverbosetrue
\caption{The four different SFHs used in the creation of the synthetic
galaxy catalogues. The fractional time runs from the big bang to the
epoch at the galaxy's redshift. The dashed lines in the top panel
indicate the blocks within which all reconstructed SFHs are
re-sampled. The early and late bursts are designed to fit entirely
within the first and last of these blocks respectively. The bolometric
luminosity of the early burst is approximately one tenth that of the
late burst.}
\label{input_SFHs}
\end{figure}

\begin{figure*}
\epsfxsize=14cm
{\hfill
\epsfbox{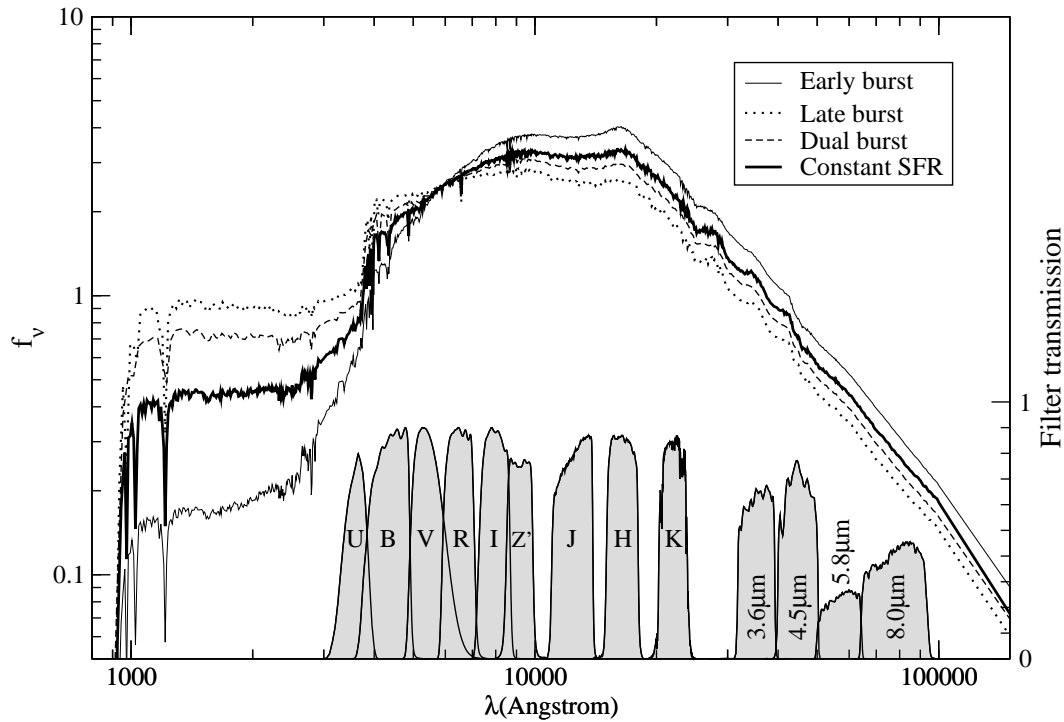}
\hfill}
\epsfverbosetrue
\caption{Synthetic SEDs corresponding to the early burst, late burst, dual
burst and constant SFR histories (see Figure \ref{input_SFHs}) for a
galaxy at $z=0$ and with $Z=0.1Z_{\odot}$, $A_V=0$. SEDs are plotted
normalised to the same $R$ band flux. The filter transmission
efficiency is shown for comparison and is correctly scaled. The total
throughput in each passband is given by scaling all filters by an
additional global system efficiency of 70\% (see text).  Left ordinate
is plotted on a log scale and applies to the SEDs, right ordinate is
linear and applies to the filter curves.}
\label{seds_and_filters}
\end{figure*}

Four different SFH types were considered. These were chosen to establish
how the reconstruction fares with the presence/absence of early and/or
late star formation activity.  The four SFH types are:
\begin{itemize}

\item {\em Early burst} - The early burst SFH starts with a high SFR
from the moment the galaxy is born followed by an exponential
decay. After approximately 40\% of the galaxy's age, the decay ceases
leaving a small SFR that remains constant for the remainder of the
galaxy's history.

\item {\em Late burst} - This SFH has a small
constant SFR from birth up until approximately 90\% of the galaxy's
age. At this point, it undergoes an instantaneous burst which
exponentially decays back to the small constant SFR the galaxy
experienced prior to the burst.

\item {\em Dual burst} - This is the early burst SFH with the last
10\% of the history replaced with the late busrst SFH.

\item {\em Constant SFR} - The SFR is constant throughout the 
entire history for this SFH.

\end{itemize}
The different SFHs are plotted in Figure \ref{input_SFHs} with a SFR
scale that corresponds to the creation of 1M$_\odot$ over the history
of the galaxy.  Absolute SFRs for each galaxy are determined by
normalising to the absolute $R$ band magnitude as described below. The
early and late bursts are designed to fit entirely within their
respective early and late SFH blocks used for re-binning in Section
\ref{sec_sims} (see Figure \ref{input_SFHs}). Although the early burst
creates approximately four times the stellar mass created in the late
burst, the bolometric luminosity of the late burst is ten times that
of the early burst.  Figure \ref{seds_and_filters} plots the SED
corresponding to each SFH type.

For every SFH type, four galaxy catalogues were generated, each 
using one of the following filtersets:
\begin{itemize}

\item[$\bullet$] {\em Full set} -- $U$, $B$, $V$, $R$, $I$, $Z'$, $J$,
$H$, $K$, $3.6\mu$m, $4.5\mu$m, $5.8\mu$m and $8\mu$m. The full set
contains all 13 broad band filters considered in this paper. The last
four bands are those of the Infra Red Array Camera (IRAC) on board the
{\sl Spitzer Space Telescope}.

\item[$\bullet$] {\em Half set} -- $B$, $R$, $I$, $J$, $K$, $3.6\mu$m
and $4.5\mu$m. The half set spans a slightly narrower range of
wavelengths than that spanned by the full set and contains half the
number of filters. This set also omits the 5.8 and 8$\mu$m IRAC
bands which are in practice dominated by dust and PAHs.

\item[$\bullet$] {\em Optical set} -- $U$, $V$, $R$, $I$, $Z'$. This
set is included to match the set of filters used by the SDSS.

\item[$\bullet$] {\em Infra-red set} -- $Z'$, $J$, $H$, $K$, $3.6\mu$m
and $4.5\mu$m. This set is purely to assess how well infra-red data
fares without optical band photometry.

\end{itemize}
The filter transmission curves are plotted in Figure
\ref{seds_and_filters} for comparison with the four different SFH type
SEDs.

Each of the 16 catalogues was populated with 1000 galaxies with random
redshifts, metallicities and extinctions. For each galaxy, 
apparent magnitudes were generated following these steps:
\begin{itemize}

\item[1)] Assign a random absolute $R$ band magnitude distributed
according to the $R$ band luminosity function of \citet{wolf03}
described by a Schechter function \citep{schechter76} with parameters
$M_*=-20.70+5{\rm lg \,h_0}$, $\alpha=-1.60$. 

\item[2)] Assign a random redshift drawn from the probability
distribution function $z \,\, {\rm exp} \, (-z^2/4)$ within the range
$0<z<6$.

\item[3)] Assign a random extinction drawn from a uniform
distribution within the range $0<A_V<3$.

\item[4)] Assign a random metallicity from a uniform logarithmic
distribution within the range $0.005 < Z/Z_{\odot} < 2.5$. (Note that
this work assumes mono-metallic SFHs, i.e., $Z$ is held constant over
the galaxy's entire history. See Section \ref{sec_summary}). Linear
interpolation in log($Z$) between the discrete metallicity library
SEDs ensures a continuous distribution in $Z$.

\item[5)] Compute the apparent $R$ band magnitude using $z$, 
the absolute $R$ band magnitude from step 1) and
the K-correction from the appropriate synthetic SED.

\item[6)] Compute fluxes in all passbands using the appropriate
redshifted, reddened but arbitrarily scaled synthetic SED.

\item[7)] Normalise each passband flux by the factor needed to scale
the $R$ band flux computed in step 6) to the apparent $R$ band
magnitude computed in step 5). Fluxes at this point are in
units of photons/s/m$^{2}$.

\item[8)] Assuming a telescope collecting area of 64m$^2$ for filters
$U$ to $K$ and 0.6m$^2$ for the four IRAC bands, an integration time
of 1800s per filter and an overall system efficiency of 70\% in all
filters, compute Poisson errors for each flux. 

\item[9)] Scatter fluxes by their errors computed in step eight then
convert the resulting fluxes and their errors to AB mags.

\end{itemize}
Once the photometry is computed for a given source in this way, the
number of filters with non-detections, defined by a flux significance
of $<10\sigma$, are counted. Sources that are not detected in at least
70\% or five (whichever is larger) of the filters contained within the
set are rejected. Sampling continues in this way until 1000 objects
have been generated for the catalogue. The 1800s exposure per filter
and telescope collecting area assumed in step eight above correspond
to the following 10$\sigma$ magnitude sensitivity limits: 26.5, 25.9,
25.6, 25.0, 24.8, 23.9, 22.8, 22.2, 22.6, 24.1, 23.5, 21.4, 21.3 in
$U$, $B$, $V$, $R$, $I$, $Z'$, $J$, $H$, $K$, $3.6\mu$m, $4.5\mu$m,
$5.8\mu$m and $8\mu$m respectively. Non-detections are assigned an
apparent magnitude equal to the sensitivity limit of the corresponding
filter and an error of 0.5 mag. The system efficiency assumed in step
8) applies in addition to the absolute filter transmission
efficiencies indicated in Figure \ref{seds_and_filters} (this brings
the IRAC filters to the correct total passband throughputs and
accommodates typical optical and IR camera throughputs).

\begin{figure}
\epsfxsize=8cm
{\hfill
\epsfbox{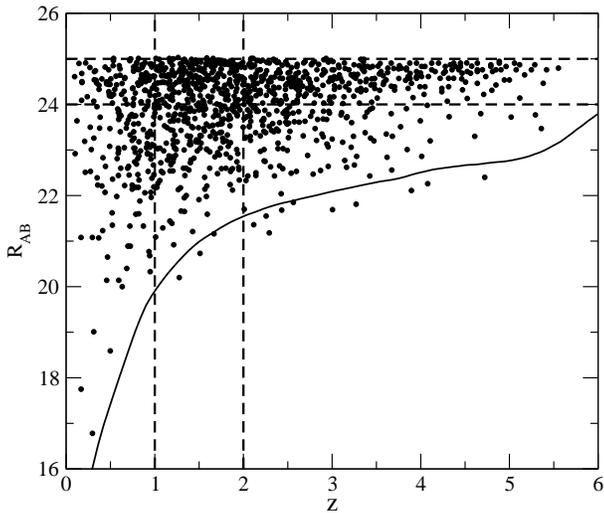}
\hfill}
\epsfverbosetrue
\caption{An example of the variation of apparent $R$ band magnitude
with redshift for all objects in one of the early burst
catalogues. The continuous line shows how $R$ varies with redshift for
a M=-22.0 early burst galaxy with $Z=0.1Z_{\odot}$ and $A_V$=0. The
dashed lines indicate bins within which objects were selected for the
analyses of Section \ref{sec_app_to_all}. The magnitude bin $24<R<25$
selects objects with an approximately constant photometric S/N over as
large a range in redshift as possible, whilst the redshift bin $1<z<2$
optimises both the number of objects and their S/N range.}
\label{R_vs_z_early_burst}
\end{figure}

In Section \ref{sec_app_to_all} the effect of photometric S/N and
redshift on the reconstructed SFHs and stellar masses is
investigated. Two catalogue sub-sets were therefore defined to achieve
this.  To test dependency on S/N with as little variation in redshift
as possible, sources within $1<z<2$ were selected. To test dependency
on as large a range in redshift as possible at approximately the same
S/N, sources were selected within $24<R<25$.  These sub-sets are shown
in Figure \ref{R_vs_z_early_burst} where the apparent $R$ band
magnitude is plotted against $z$ for 1000 sources generated using the
early burst SFH.

\section{Simulation results}
\label{sec_sims}

This section discusses application of the SFH reconstruction method to
synthetic catalogues to assess its performance.  An initial
demonstration of setting the optimal regularisation is given in
Section \ref{sec_reg_effect} before applying the method to the full
range of catalogues in Section \ref{sec_app_to_all}.

\subsection{The effect of regularisation}
\label{sec_reg_effect}

The effect of regularisation is demonstrated with an example.  Using
the late burst SFH, synthetic photometry was generated in the full
filterset for a galaxy at $z=1$ with absolute $R$ band magnitude
$M_R=-20$, $A_V=0$ and $Z=0.1Z_{\odot}$. The resulting stellar mass of
the galaxy was $9.7\times 10^9 \, {\rm M}_{\odot}$.  The SFH
reconstruction method was then applied for different degrees of
regularisation.  In each case, the SFH was divided into five
exponentially spaced blocks as indicated in the top panel of Figure
\ref{input_SFHs} (see Section \ref{sec_max_proc}). For comparison with
the reconstructed SFHs, the input SFH was binned into the same five
exponentially spaced blocks.

\begin{figure}
\epsfxsize=8cm
{\hfill
\epsfbox{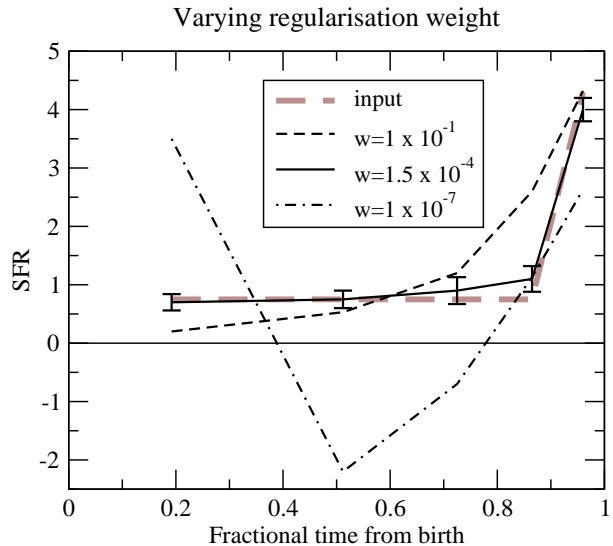}
\hfill}
\epsfverbosetrue
\caption{Demonstration of the effect of different regularisation 
weights, $w$, on the reconstructed SFH. This example is based on a
synthetic source lying at $z=1$ with $Z=0.1Z_{\odot}$, $M_R=-20$ and
$A_V=0$. The reconstruction uses the full set of 13 filters.  The
input SFH is the late burst model shown here by the thick grey dashed
line, binned into five SFH blocks. The optimal regularisation weight
found by maximising the Bayesian evidence produces the most accurate
reconstructed SFH (continuous line). Under-regularisation (dot-dashed
line) results in a very inaccurate SFH reconstruction whereas
over-regularisation (thin dashed line) smooths the SFH too heavily.
For clarity, the standard errors returned by equation
(\ref{eq_sfr_errors}) are shown only for the optimally regularised
case. In all cases, the points are placed at the SFH block centres.}
\label{example_reg}
\end{figure}

Figure \ref{example_reg} shows how accurately the input late burst SFH
was reconstructed with three different values of the regularisation
weight, $w$. One of these values is the optimal weight,
$w=1.5\times 10^{-4}$, as determined by the maximal evidence,
whilst the remaining two were set higher and lower than this by $\sim
3$ dex. In the figure, the input binned SFH is shown by the heavy
dashed line.  Clearly, the optimal regularisation weight gives the
most accurate reconstruction.  Over-regularisation smooths the SFH too
heavily, leading to a biased reconstructed SFH. Conversely,
under-regularisation gives rise to a catastrophic failure, with the
SFH ringing violently about the input SFH.

The exercise also serves to demonstrate that the reconstructed stellar
mass (computed using equation \ref{eq_stellar_mass}) depends on $w$.
Comparing with the stellar mass of the input galaxy of $9.7\times 10^9
\, {\rm M}_{\odot}$, the optimally regularised case recovered a mass
of $(9.9\pm0.3)\times 10^9 \, {\rm M}_{\odot}$, the under-regularised
case recovered $(1.61\pm0.06)\times 10^{10} \, {\rm M}_{\odot}$ and the
over-regularised case recovered $(8.9\pm0.2)\times 10^9 \, {\rm
M}_{\odot}$. A sub-optimal regularisation weight can therefore bias
the reconstructed mass. 

As stated previously, the actual number of SFH blocks is always higher
than the effective number of blocks when regularising due to the
smoothness constraints imposed on the SFH. To reiterate, this is why
the evidence should be the statistic used to rank models rather than
$\chi^2$. These constraints increase the covariance between pairs of
SFH blocks although the effect is counteracted by the evidence which
selects fewer SFH blocks (and hence less covariant solutions) when the
data do not support a high resolution SFH. Inspection of many
realisations of the covariance matrix (excluding failed
reconstructions -- see next section) indicates that highly covariant
solutions do not occur. A further observation is that the early blocks
are always more covariant than the later blocks.

\subsection{Application to the full suite of catalogues}
\label{sec_app_to_all}

The SFH reconstruction method was applied to the full suite of
catalogues. An assessment was made of how the accuracy of the
reconstructed SFH depends on the number of filters, the wavelength
range spanned by the filterset, the S/N of the photometry, the
presence and/or absence of early and/or late star formation activity
and redshift. For each combination of these variables, a synthetic
catalogue was generated, comprising 1000 galaxies adhering to the
ranges in $z$, $A_V$, $Z$ and absolute magnitude given in Section
\ref{sec_synthetic_cats}. For every object in each case, the SFH was
reconstructed following the procedure outlined in Section
\ref{sec_max_proc}, maximising the evidence by varying the
regularisation weight, number of SFH blocks and metallicity.

\begin{figure*}
\epsfxsize=17.8cm
{\hfill
\epsfbox{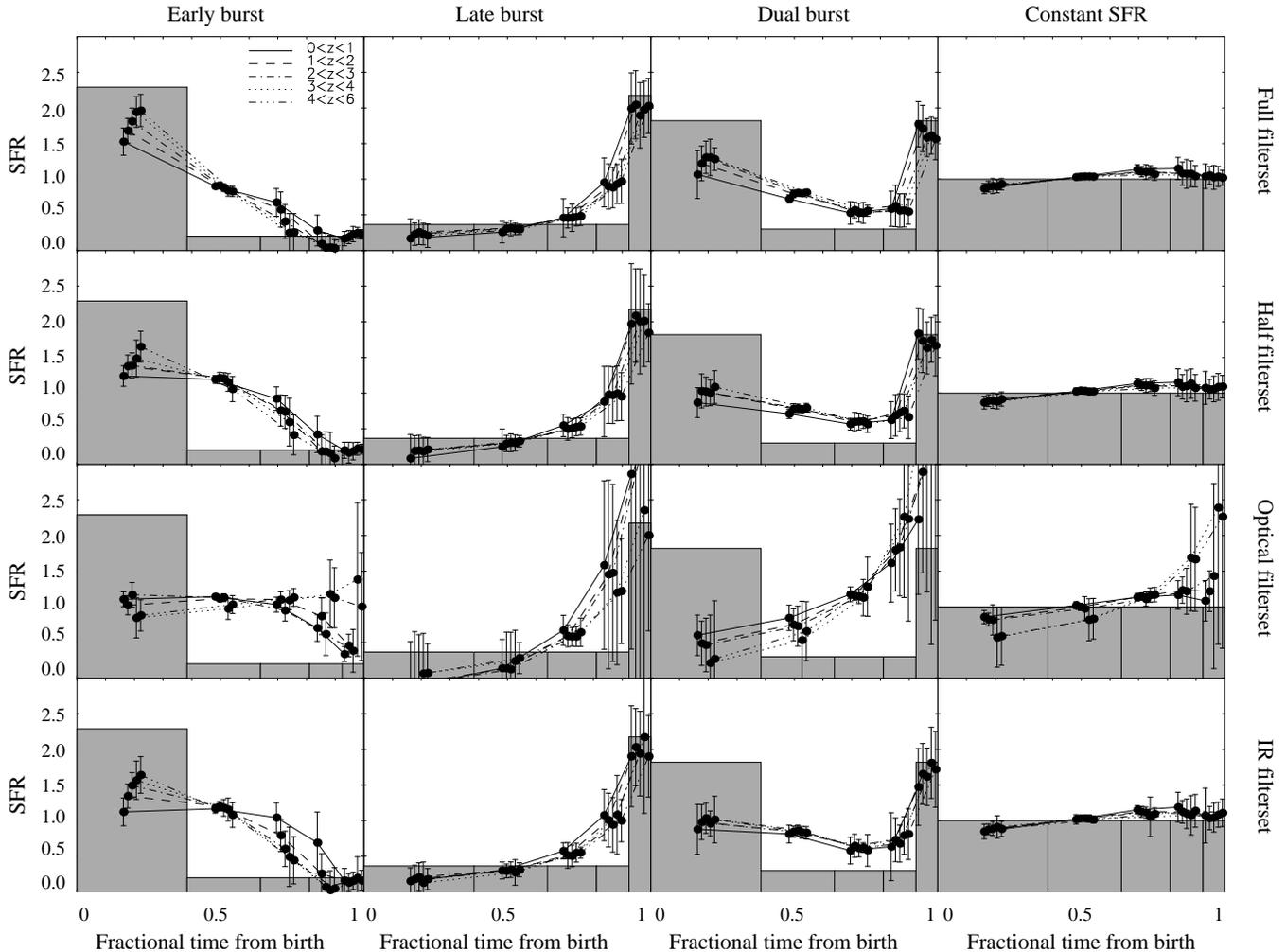}
\hfill}
\epsfverbosetrue
\caption{SFH reconstruction binned by redshift as labelled.  SFH type
is separated by column and filterset by row.  Reconstructed SFHs are
shown by the data points and lines (staggered for clarity) and apply
to objects selected by $24<R<25$. Error bars show the standard
deviation of objects in the redshift bin. Grey shaded histograms are
the binned input SFHs.}
\label{sfh_recon1}
\end{figure*}

\begin{figure}
\epsfxsize=7cm
{\hfill
\epsfbox{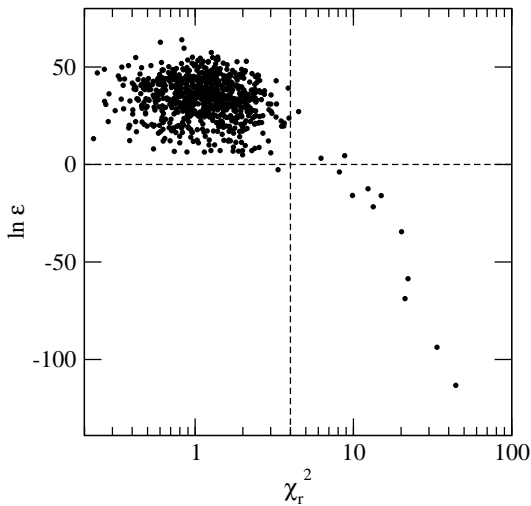}
\hfill}
\epsfverbosetrue
\caption{Distribution of 1000 reconstructions in the plane spanned by
log-evidence and reduced $\chi^2$ for the early burst SFH and full
filterset. Catastrophic failures lie in the tail extending to
low $\epsilon$ and high $\chi_r^2$ and are removed in all analyses
in this paper using the limits ln$\,\epsilon>0$
and $\chi_r^2<4$ indicated by the dashed lines.}
\label{lnE_vs_chisq}
\end{figure}

The results show that approximately 1\% of reconstructions completely
fail to recover the input SFH or galaxy parameters. The size of this
fraction is independent of SFH type or filterset. These catastrophic
failures occur either when the maximisation becomes stuck at an
incorrect local maximum or when the maximisation fails to
converge. Fortunately, these cases are easily identified by their very
small evidence and large $\chi^2$. Figure \ref{lnE_vs_chisq} shows the
distribution of sources in the ln$\,\epsilon$, $\chi^2$ plane for
the early burst SFH and full filterset reconstruction (see next
section).  The catastrophic failures form the long tail extending to
low $\epsilon$ and high $\chi^2$ and can be discounted by retaining
only objects with ln$\,\epsilon>0$ and $\chi_r^2<4$. In all analyses
hereafter, this cut has been applied. The figure also serves to
illustrate that there is not a clear relationship between the evidence
and $\chi^2$, i.e., minimising $\chi^2$ is by no means equivalent
to minimising $- {\rm ln}\,\epsilon$.

\subsubsection{Dependence on filterset and SFH type}

Figure \ref{sfh_recon1} shows how the method performs as a function of
SFH type, filterset and redshift. Each panel corresponds to a
different combination of SFH type and filterset and in every panel,
the average reconstructed SFH and its standard deviation is plotted
for sources in five different redshift bins: $0<z<1$, $1<z<2$,
$2<z<3$, $3<z<4$ and $4<z<6$. To allow for variation in the number of
preferred SFH blocks from source to source, each reconstructed SFH was
finely sampled with a small fixed time step then re-binned to a
common five-block SFH.  An effect of the re-binning is to smear the
reconstructed SFHs slightly, particularly when re-binning from a lower
number of blocks.  However, comparing with SFHs averaged over only
those sources preferring five bins shows that this effect is
relatively minor with no more than five per cent of the total stellar
mass being smeared between any pair of bins in all cases.

The results plotted in Figure \ref{sfh_recon1} illustrate that the SFH
type and filterset have a strong influence on the accuracy with which
the input SFH can be recovered. In terms of the filters, the full set
unsurprisingly performs best.  However, a mildly surprising find is
that the half set gives very similar average SFHs, albeit with $\sim
30\%$ larger scatter on average. Clearly, the wavelength range spanned
by the filterset is the important factor, rather than the existence of
an extra six intermediate photometric points provided by the full set.
Furthermore, the IR end of the filterset is more important than the
optical end, as indicated by the bottom two rows of Figure
\ref{sfh_recon1}. The optical SDSS-like set performs poorly,
significantly worse than the IR set. Only in the specific case of the
late burst does the SFH reconstructed using optical photometry
consistently resemble the input SFH, but this can not be reliably
distinguished from the other cases.

In terms of the SFH type, the late burst and constant SFHs are
reconstructed the most faithfully although the late burst is smeared
slightly towards earlier times. The early burst reconstructions are
more strongly smeared over the first few bins, giving rise to less
star formation at early times and more during their mid-history than
actually occurred. On average, $\sim 20\%$ of the stellar mass created
in the early burst is smeared into the later blocks. The stronger
smearing exhibited by the early burst is a result of its bolometric
luminosity being ten times smaller than that of the late burst.
Nevertheless, the reconstructed SFHs still prove a useful diagnostic
for the presence of early star formation activity, showing a clear
excess that declines with time to accurately reproduce the latest SFR
(with the exception of the optical filterset which fails to recover a
decline at all redshifts). The dual burst proves the most challenging
of SFH types to reconstruct. In this case, the full and half
filtersets best recover the early and late bursts, implying the
necessity of both optical and IR filters, although sources at $z<1$
have more strongly smeared SFHs. Again, this demonstrates the
importance of the IR filters.

\begin{figure*}
\epsfxsize=17.8cm
{\hfill
\epsfbox{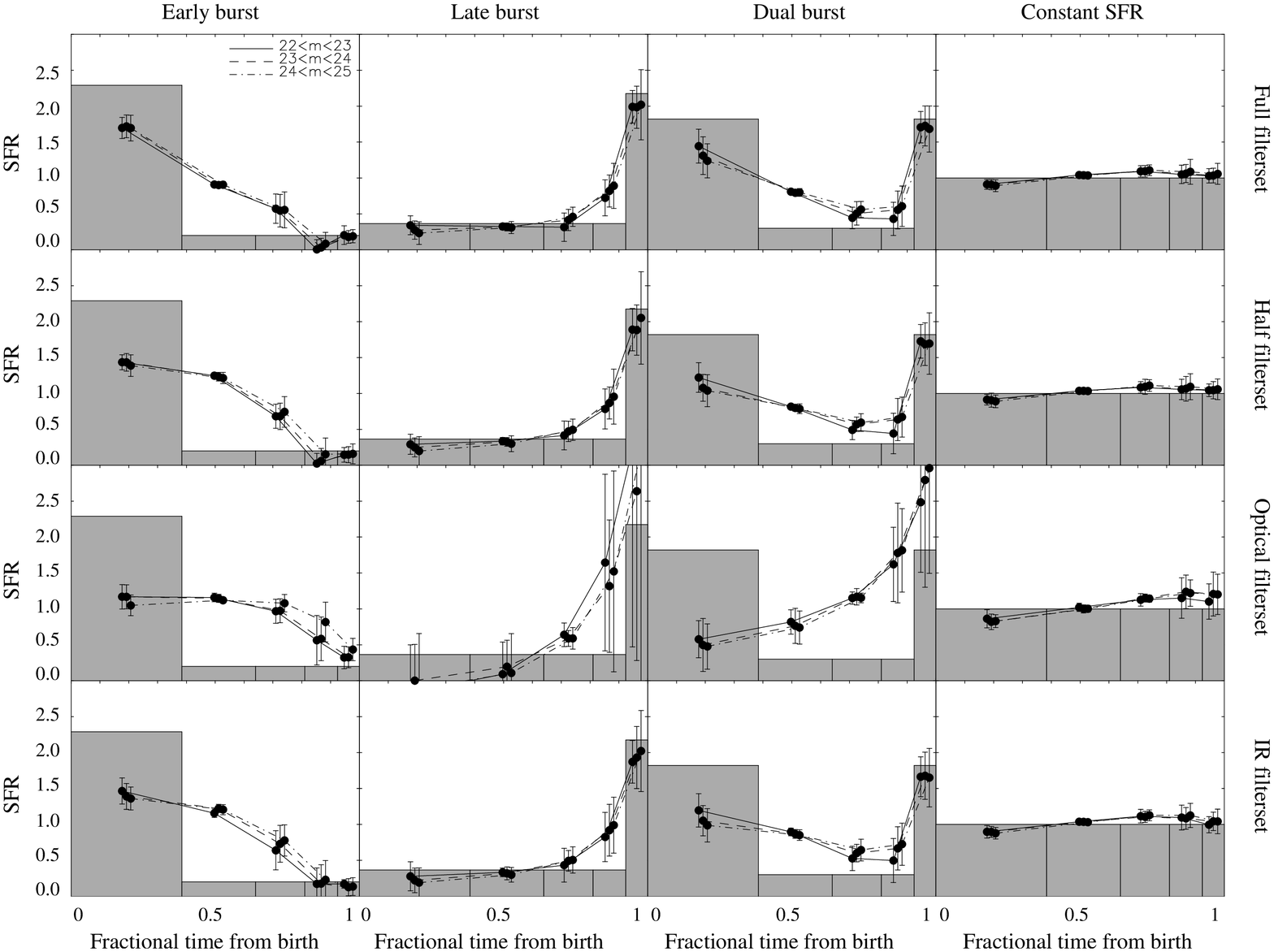}
\hfill}
\epsfverbosetrue
\caption{SFH reconstruction binned by magnitude as labelled.  SFH type
is separated by column and filterset by row.  Reconstructed SFHs are
shown by the data points and lines (staggered for clarity) and apply
to objects selected by $1<z<2$. Error bars show the standard deviation
of objects in the redshift bin. Grey shaded histograms are the binned
input SFHs.}
\label{sfh_recon2}
\end{figure*}

Note that the effect of regularisation on the average of a sample of
SFHs is twofold. A stronger regularisation weight reduces the scatter
in the sample, whilst more heavily smoothing the average SFH. This
effect can be seen to an extent by comparing the reconstructed early
burst SFH for the full and half filtersets in Figure
\ref{sfh_recon1}. The error bars on points in the first bin with the
half filterset are of equal size or smaller than the error bars of the
first bin with the full set. However, the SFHs are more heavily
smeared with the half set.

\subsubsection{Dependence on S/N and redshift}

Figure \ref{sfh_recon1} shows that in nearly all cases, the variation
in reconstructed SFHs between different redshift bins is comparable to
or less than the intrinsic SFH scatter within a given bin. Generally,
the low redshift sources (selected by say $z<2$) tend to have more
smeared SFHs than their higher redshift equivalents. This is
consistent with the fact that at $z\simgreat 2$, the rest-frame UV is
redshifted into the optical wavebands where SEDs are much more
sensitive to stellar age (see Figure \ref{seds_and_filters} -- note
that the optical filterset performs worst despite this since it lacks
the SED normalisation provided by the IR filters). Furthermore, since
the SFHs in Figure \ref{sfh_recon1} are computed for sources selected
by $24<R<25$ (i.e., they have approximately the same photometric S/N),
the flux received by the IRAC filters increases with redshift,
providing more discrimination at the IR end of the SED.

Figure \ref{sfh_recon2} shows reconstructed SFHs for the different
combinations of filterset and SFH type, but this time objects are
binned by apparent magnitude. All objects are selected by $1<z<2$ to
maximise the number of objects whilst maintaining a large span in
apparent magnitude and thus S/N. As the figure shows, there is little
variation with S/N. The averaged SFHs are very similar, although
unsurprisingly, the scatter increases as the apparent magnitude falls.

As can be inferred from Figures \ref{sfh_recon1} and \ref{sfh_recon2},
the reconstructed SFH can give rise to negative SFRs. This is
especially true of the inadequate optical filterset. With the other
three filtersets, negative SFRs still occur but such cases; 1) tend
to be limited to galaxies with low S/N photometry, 2) are always
consistent with a null SFR, 3) are relatively infrequent due to the
optimal regularisation strength selected by the evidence.

\subsubsection{Recovery of stellar mass and metallicity}

\begin{figure*}
\epsfxsize=14cm
{\hfill
\epsfbox{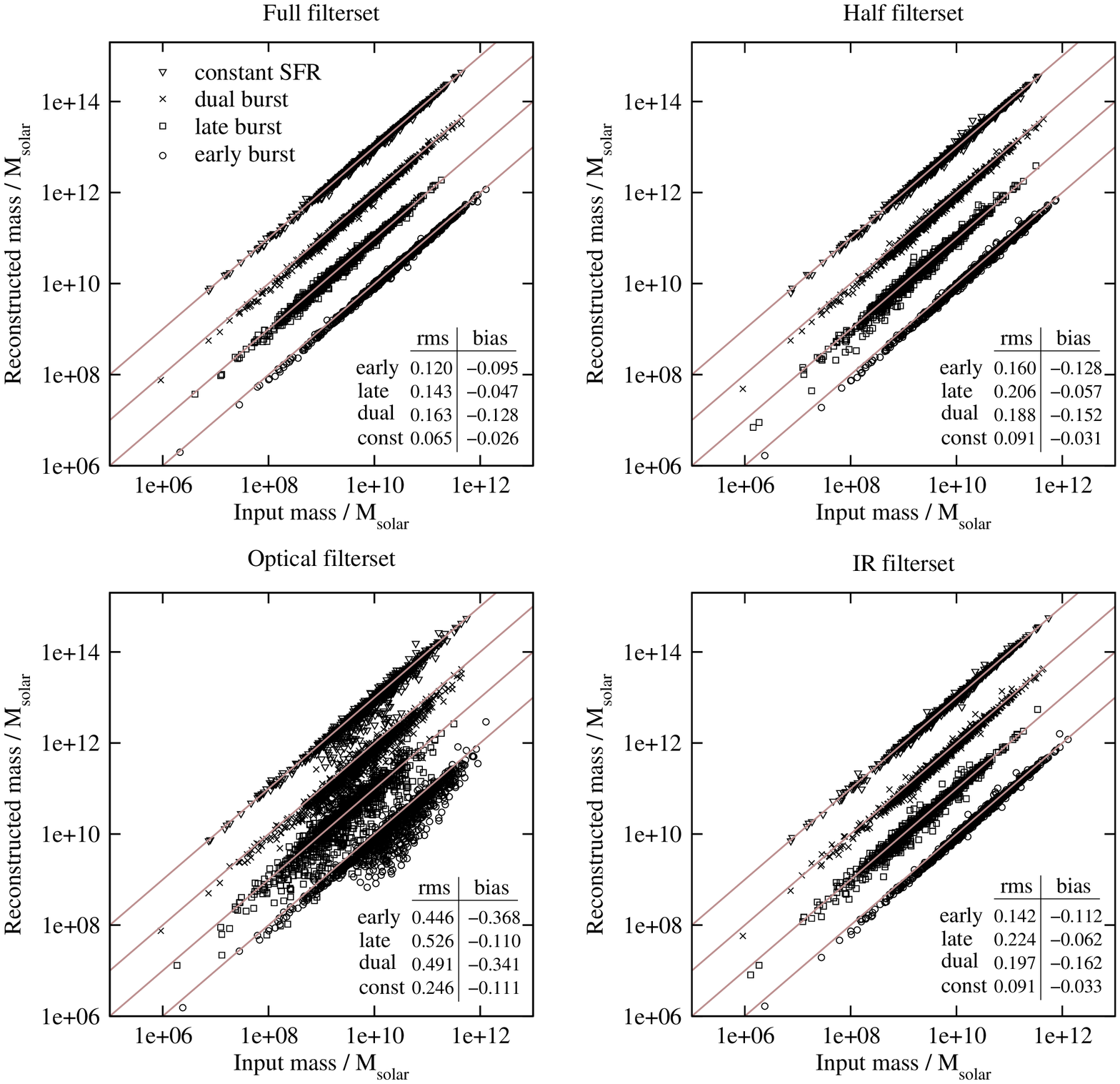}
\hfill}
\epsfverbosetrue
\caption{Accuracy of reconstructed mass. Each panel corresponds to a
different filterset as labelled. For each filterset, the
reconstructed mass is plotted against the input mass for the early
burst SFH, late burst SFH (reconstructed mass $\times 10$), dual burst
SFH (reconstructed mass $\times 100$) and constant SFR (reconstructed
mass $\times 1000$). Tables in the bottom right of each panel list the
fractional scatter $\left<({\rm M}_{\rm recon}-{\rm M}_{\rm input})^2/
{\rm M}_{\rm input}^2\right>^{1/2}$ and the bias $\left<({\rm M}_{\rm
recon}-{\rm M}_{\rm input})/ {\rm M}_{\rm input}\right>$ for the
different SFH types.}
\label{mass_cf}
\end{figure*}

Figure \ref{mass_cf} shows the recovered stellar mass as a function of
the input mass for the different combinations of SFH type and filter
set. In the lower right hand corner of each panel, a table lists the
fractional scatter $\left<({\rm M}_{\rm recon}-{\rm M}_{\rm input})^2/
{\rm M}_{\rm input}^2\right>^{1/2}$ and the bias $\left<({\rm M}_{\rm
recon}-{\rm M}_{\rm input})/ {\rm M}_{\rm input}\right>$ for each SFH
type.

As expected, the full filterset recovers the stellar mass most
accurately (smallest bias) and with the least scatter. However, all
cases show a negative bias such that the recovered mass is on average
less than the input mass. For the full filterset, this bias ranges
from $\sim 3\%$ for the constant SFR to $\sim 13\%$ for the dual burst
SFH.  The largest bias of $\sim 40\%$ occurs with the early burst SFH
and optical filterset.  However, in all cases, the bias is less than
the fractional scatter.

Compared with the full filterset reconstructions, the half set again 
performs very well given the reduction from 13 filters to seven. The
fractional scatter of the half set is higher than that of the full
set by $\sim 25\%$ on average. Similarly, the IR filterset results in
an increased fractional scatter of only $\sim 30\%$ compared to the 
full set on average. The optical filterset gives a significantly
larger scatter of around four times that of the full set or three times
the IR set on average, confirming the well known fact that IR photometry
is essential for the accurate measurement of stellar mass.

In terms of the dependence of mass recovery on SFH type, the constant
SFR masses show the smallest bias, closely followed by those of the
late burst (although the late burst gives rise to significantly more
scatter). The early burst masses tend to be more accurately
reconstructed than the late burst or dual burst masses, especially in
the case of the IR filterset where they are recovered almost as
accurately as the full filterset case. This demonstrates the
importance of IR filters for measuring stellar mass created in early
bursts.

\begin{figure*}
\epsfxsize=16cm
{\hfill
\epsfbox{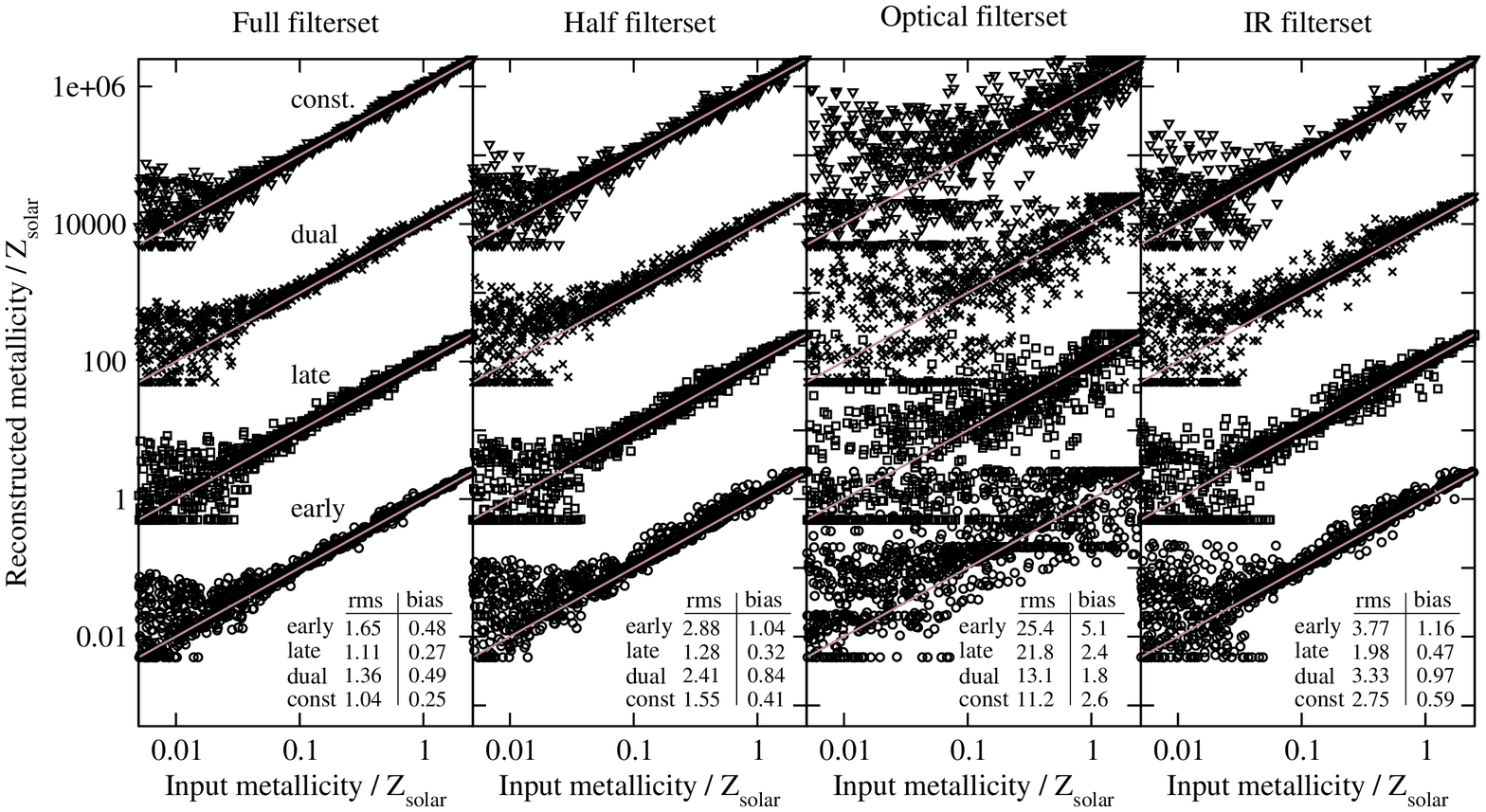}
\hfill}
\epsfverbosetrue
\caption{Accuracy of reconstructed metallicity. Each panel corresponds 
to a different filterset as labelled. For each filterset, the
recovered metallicity is plotted against the input metallicity for the
early burst SFH, late burst SFH (reconstructed $Z$ $\times 10^2$), dual
burst SFH (reconstructed $Z$ $\times 10^4$) and constant SFR
(reconstructed $Z$ $\times 10^6$). Tables in the bottom right of each
panel list the fractional scatter $\left<(Z_{\rm recon}-
Z_{\rm input})^2/ Z_{\rm input}^2\right>^{1/2}$ and the bias
$\left<(Z_{\rm recon}-Z_{\rm input})/ Z_{\rm
input}\right>$ for the different SFH types.}
\label{Z_cf}
\end{figure*}

Figure \ref{Z_cf} plots the recovered metallicity as a function of the
input metallicity for all SFH types and filtersets. The scatter in the
recovered metallicity, particularly at low $Z$ ($<0.1 Z_{\odot}$), is
larger than the scatter seen in the reconstructed mass but the global
trends are essentially the same. The full filterset recovers
metallicity most accurately, the half filterset and IR filterset
having a scatter larger by $\sim 60\%$ and $\sim 120\%$ respectively on
average. The very large scatter exhibited by the optical filterset
demonstrates that recovery of metallicity without IR filters is
extremely unreliable. In all cases, the recovered metallicity is
larger than the input value, although similar to the recovered mass,
this bias is always significantly lower than the scatter.

\subsubsection{SFH resolution}

In the previous sections, SFHs were re-binned to bring them to a
common resolution of five blocks to enable comparison between
reconstructions. In this section, dependency of the reconstructed SFH
resolution (i.e., number of blocks, $N_{\rm block}$) on data quality,
SFH type and filterset is considered.

\begin{figure}
\epsfxsize=8.2cm
{\hfill
\epsfbox{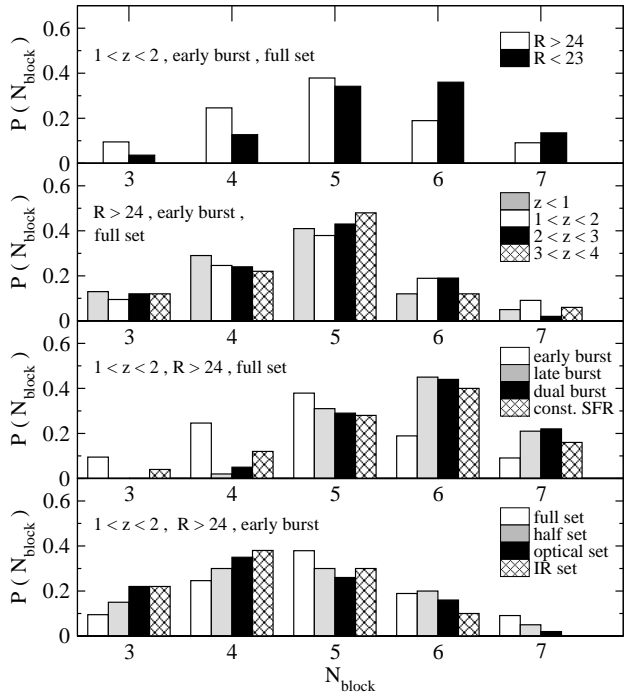}
\hfill}
\epsfverbosetrue
\caption{Distribution of the optimal number of SFH blocks, $N_{\rm
block}$, chosen by the Bayesian evidence for different redshift and
magnitude selections, SFHs and filtersets. The reference selection
shown by the unshaded histogram in each panel satisfies the criteria
$1<z<2$ and $R>24$ with the full filterset and early burst SFH.}
\label{nblock_hist}
\end{figure}

Figure \ref{nblock_hist} shows how the distribution of $N_{\rm block}$
varies as the data vary. The top panel shows that higher S/N data
allow a higher SFH resolution, sources selected by $R<23$ preferring
five to six blocks on average, compared with $R>24$ sources preferring
an average of four to five blocks. The panel second from top shows how
the resolution varies as a function of redshift for sources of
approximately constant S/N ($R>24$). The differences are not
significant, with sources across all redshifts preferring five bins on
average.

The third panel from top in Figure \ref{nblock_hist} shows how the SFH
resolution depends on SFH type. In this case, there are more
significant differences. The late burst, dual burst and constant SFR
histories allow a higher resolution of six blocks on average, compared
to the early burst of five. Finally, the bottom panel shows the
dependence of resolution on filter set. Unsurprisingly, the full set
allows the highest resolution on average, with the majority of
galaxies preferring four or five SFH blocks. In comparison, the
distribution in resolution of the reduced filtersets is skewed to
lower numbers of SFH blocks, particularly the IR set.

\begin{figure}
\epsfxsize=8.0cm
{\hfill
\epsfbox{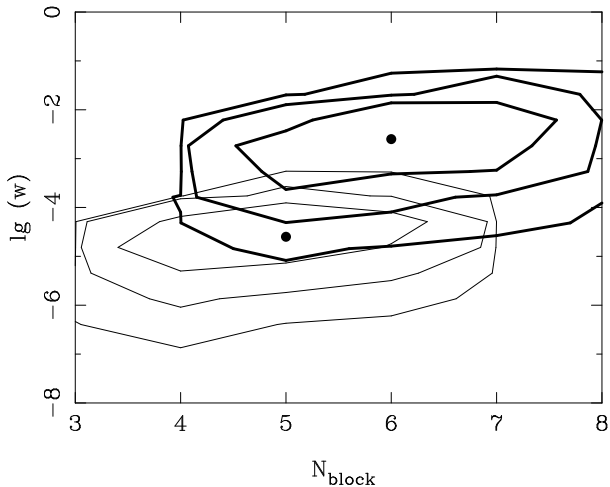}
\hfill}
\epsfverbosetrue
\caption{Confidence limits on regularisation weight, $w$,
and number of SFH blocks, $N_{\rm block}$ for a $z=1$,
$Z=0.1Z_{\odot}$, $M_R=-18$ and $A_V=0$ galaxy generated using the
early burst SFH (thin contours) and late burst SFH(thick contours).
Contours are computed from the evidence and correspond to
68\%, 95.4\% and 99.7\% confidence levels.}
\label{ev_contours}
\end{figure}

Clearly, there is a degeneracy between the SFH resolution and the
regularisation weight, since a higher level of regularisation acts to
smooth the SFH, effectively reducing its resolution. In Figure
\ref{ev_contours}, two example confidence regions are shown in the
plane spanned by regularisation weight and $N_{\rm block}$ computed
from the Bayesian evidence. The heavy contours correspond to the late
burst SFH and the thin contours the early burst SFH for a $z=1$,
$Z=0.1Z_{\odot}$, $M_R=-18$ and $A_V=0$ galaxy. The inclination of
both contours shows that this degeneracy does indeed exist. However,
the degeneracy is weak and therefore locating the maximum in the
evidence distribution is relatively straightforward.

Figure \ref{nblock_hist} illustrates that the number of SFH blocks
that can be recovered on average is comparable to the typical number
recovered by \citet{tojeiro07} from optical spectra. However, there
are two major differences with the present study that make this an
unfair comparison. These are that mono-metallic populations are
considered and that filtersets extend to the IR. Increasing the
number of parameters to describe a time-varying metallicity will
reduce the number of SFH blocks that can be recovered (see Section
\ref{sec_summary}). Similarly, the IR filters provide extra
constraints on the SFH, allowing reconstruction at a slightly higher
resolution. As Figure \ref{nblock_hist} shows (for the early burst,
but this applies generally), the full filterset recovers more SFH
blocks on average than the optical set even though the optical set is
similar but lacking the IR bands.

\section{Summary}
\label{sec_summary}

The primary aim of this study has been to assess reconstruction of
discretised SFHs using a new method applied to multi-band photometric
data. Although not tested in this paper, the method can also be
applied to spectroscopic data as well as a mixture of both
spectroscopic and multi-band data. 

The method differs from existing methods by maximising the Bayesian
evidence instead of minimising $\chi^2$ (or maximising the posterior
probability).  For regularised solutions, the evidence gives the
unbiased relative probability of the fit between different model
parameterisations. This is unlike the $\chi^2$ statistic which suffers
from an ambiguous number of degrees of freedom that changes between
parameterisations when regularisation is applied. 

This work has demonstrated that the evidence allows the data to
correctly and simultaneously set the optimal regularisation strength
and the appropriate number of blocks in the reconstructed SFH.
Although negative SFRs can arise, the optimal level of regularisation
ensures that the fraction of such cases is low. Negative SFRs are
limited mainly to galaxies with low photometric S/N and inadequate
filter sets (e.g., the optical set considered in this work).  Provided
the filter set is adequate, negative SFRs are always consistent with a
null SFR. This approach may be preferable to schemes that enforce
positive SFRs. Enforcing positivity not only risks artificial ringing
and biasing in the reconstructed SFH, it also hides problems that give
rise to negative SFRs.

Application of the method to a range of synthetic galaxy catalogues
generated with varying passband sets and SFHs demonstrates that use
of multi-band data in constraining SFHs is feasible along with certain
caveats. The scatter seen in the SFHs reconstructed in this work shows
that occasional significant inaccuracies can occur even with a
comprehensive filterset that extends up to near-IR and mid-IR
wavelengths. Therefore, interpretation of SFHs recovered from solely
multi-band photometry on a galaxy by galaxy basis should be conducted
with some caution. The mean SFH of a sample of galaxies is therefore a
more reliable quantity in order to average out uncertainties although
this study indicates that averaging over only four galaxies readily
allows a late burst to be distinguished from an early burst.  In
comparison, studies using spectroscopic data show that reliable SFHs
can be derived for individual galaxies. Nevertheless, multi-band
photometry allows reconstruction of SFHs for many times more galaxies
than spectroscopic methods for the same amount of observing time.

The most important factor governing the accuracy of the reconstructed
SFHs is the wavelength range spanned by the filterset. The results
show little difference between two filtersets that span approximately
the same wavelength range (optical to mid-IR) despite one set having
half the number of filters of the other. Conversely, SFHs based on
only purely optical photometry are completely unreliable, it being
impossible to distinguish any of the input SFHs investigated.  A
filterset consisting of only near and mid IR filters ($Z'$ --
4.5$\mu$m) allows recovery of SFHs to within a comparable accuracy to
that recovered when optical filters are also included, implying that
the majority of the SFH constraints are provided by near and mid-IR
data (for the SFHs tested here).

In terms of the ability of multi-band photometry to constrain
different SFH types, the results show that apart from the case where
only optical filters are used, early bursts of star formation can be
differentiated from late bursts and both of these can be distinguished
from dual bursts and constant SFRs.  However, early bouts of star
formation activity are always artificially smeared to later times in
the reconstructed SFH compared to the input SFH. These findings apply
specifically to the SFHs considered in this work, where the early
burst gives rise to bolometric luminosity that is one tenth that of
the late burst. A quick test has revealed that a stronger early burst
is more accurately recovered with less smearing to late times. In
addition, although the dual burst SFH used here suggests that recovery
of more than two bursts would be unfeasible with the filtersets
tested, bursts with more similar bolometric luminosities can be more
readily recovered.  This was demonstrated by \citet{ocvirk06} who
showed that CSP SEDs constructed from flux normalised bursts allow a
higher SFH resolution on average than SEDs constructed from mass
normalised bursts.

The results presented in this paper have been obtained using the
\citet{bruzual03} spectral libraries. Whilst the exact values of the
numerical results quoted here will depend on the specific SED library
of choice, there are no compelling reasons to suggest that the
observed trends would not remain valid generally.

This study has considered a specific case where galaxy redshift and
extinction are known prior to reconstructing the SFH. Also,
mono-metallic stellar populations have been assumed where the
metallicity does not evolve as the galaxy ages. Clearly, the more
general problem necessitates maximising the evidence over extra
parameters. The expected effect of this is that the maximum evidence
would shift to lower SFH resolutions on average. Although generalising
to a variable redshift and extinction is a relatively small expansion
of the non-linear parameter space, incorporating a time-varying
metallicity in addition results in a significantly larger and more
complex non-linear parameter space. This increases the time required
to locate the maximum evidence and increases the risk of becoming
trapped at a local maximum.

However, there are two small reprieves.  The first is that the
metallicity history can be regularised in a similar manner to the SFH,
smoothing the evidence surface and therefore easing maximisation. The
second exploits SED libraries with discrete metallicities. As shown by
\cite{tojeiro07}, finding the optimal metallicity within the range
spanned by two tabulated values of metallicity is also a linear
problem which can be directly combined with the linear inversion of
the SFH. In this way, optimising the metallicity for each SFH block
reduces to searching a smaller number of discrete values. A full
investigation of the general case will be presented in forthcoming
work.

\appendix
\section{Ranking models by Bayesian evidence}
\label{sec_app_evidence}

In this appendix, the theory of \citet{suyu06} is applied to the
present problem.  Using the notation in Section
\ref{sec_most_prob_SFH}, Bayes' theorem states that the posterior
probability of the parameters $\mathbf{a}$ given the data
$\mathbf{d}$, model $\mathbf{G}$ and regularisation
($\mathbf{H},w$) can be expressed as
\be
\label{eq_bayes1}
P(\mathbf{a} | \mathbf{d},\mathbf{G},\mathbf{H},w)=
\frac{P(\mathbf{d} | \mathbf{a},\mathbf{G})
P(\mathbf{a} | \mathbf{H},w) }
{P(\mathbf{d} | \mathbf{G},\mathbf{H},w)} \, .
\ee
Here, $P(\mathbf{d} | \mathbf{a},\mathbf{G})$ is the {\em likelihood}
which gives the probability of the data given the model parameters,
$P(\mathbf{a} | \mathbf{H},w)$ is the {\em prior} which
forces an a priori assumption on the parameters $\mathbf{a}$
given some regularisation model and
$P(\mathbf{d} | \mathbf{G},\mathbf{H},w)$ is a normalisation
term called the {\em evidence}.

In the first level of inference where the most likely parameters
$\mathbf{a}$ are determined, the evidence is constant and therefore
plays no part. The evidence becomes relevant in the higher levels.
For example, in the second level where one wishes to find
the optimal regularisation weight, $w$, Bayes' theorem shows
that the following posterior probability must be maximised:
\be
P(w | \mathbf{d},\mathbf{G},\mathbf{H})=
\frac{P(\mathbf{d} | \mathbf{G},\mathbf{H},w) P(w) }
{P(\mathbf{d} | \mathbf{G},\mathbf{H})} \, .
\ee
The first term in the numerator of this equation is the
evidence in equation (\ref{eq_bayes1}).

In the third level of inference where different models and
regularisation types are ranked, the posterior probability, again
using Bayes' theorem can be written 
\be
\label{eq_bayes2}
P(\mathbf{G},\mathbf{H} | \mathbf{d}) \propto
P(\mathbf{d} | \mathbf{G},\mathbf{H})P(\mathbf{G},\mathbf{H}) \, .
\ee
With a flat prior $P(\mathbf{G},\mathbf{H})$,
the likelihood $P(\mathbf{d} | \mathbf{G},\mathbf{H})$ can
be used to rank the data. The likelihood can be written
\be
P(\mathbf{d} | \mathbf{G},\mathbf{H}) = 
\int P(\mathbf{d} | \mathbf{G},\mathbf{H},w)
P(w) {\rm d} w
\ee
where, again, $P(\mathbf{d} | \mathbf{G},\mathbf{H},w)$ is the
evidence in equation (\ref{eq_bayes1}). As \citet{suyu06} discuss, a
reasonable approximation is to treat $P(w)$ as a delta function
centred on the optimal regularisation weight, $\hat{w}$, since
$\hat{w}$ has a well defined value estimable from the data. In this
case, with a flat prior $P(\mathbf{G},\mathbf{H})$, the model ranking
posterior probability given by equation (\ref{eq_bayes2}) is simply
equal to the value of the evidence at $\hat{w}$.  The most
probable model ($\mathbf{G},\mathbf{H}$) is therefore found by
maximising the evidence at $\hat{w}$.

\begin{flushleft}
{\bf Acknowledgements}
\end{flushleft}

SD is supported by the Particle Physics and Astronomy Research Council
and thanks Steve Eales and Luca Cortese for helpful discussion.

%%%%%%%%%%%%%%%%%%%%%%%%%%%%%%%%%%%%%%%%%%%%%%%%%%%%%%%%%%%%

{}

\label{lastpage}

\end{document}